# Robust Power Flow and Three-Phase Power Flow Analyses

Amritanshu Pandey[1], *Graduate Student Member, IEEE*, Marko Jereminov[1], *Graduate Student Member, IEEE,* Martin R. Wagner[1], *Graduate Student Member, IEEE,* David M. Bromberg[1], *Member, IEEE,* Gabriela Hug[2], *Senior Member, IEEE* and Larry Pileggi[1], *Fellow, IEEE*

*Abstract*—Robust simulation is essential for reliable operation and planning of transmission and distribution power grids. At present, disparate methods exist for steady-state analysis of the transmission (power flow) and distribution power grid (three-phase power flow). Due to the non-linear nature of the problem, it is difficult for alternating current (AC) power flow and three-phase power flow analyses to ensure convergence to the correct physical solution, particularly from arbitrary initial conditions, or when evaluating a change (e.g. contingency) in the grid. In this paper, we describe our equivalent circuit formulation approach with current and voltage variables that models both the positive sequence network of the transmission grid and three-phase network of the distribution grid without loss of generality. The proposed circuit models and formalism enable the extension and application of circuit simulation techniques to solve for the steady-state solution with excellent robustness of convergence. Examples for positive sequence transmission and three-phase distribution systems, including actual 75k+ nodes Eastern Interconnection transmission test cases and 8k+ nodes taxonomy distribution test cases, are solved from arbitrary initial guesses to demonstrate the efficacy of our approach.

## I. Introduction

An interconnected electric grid is a network of synchronized power providers and consumers that are connected via transmission and distribution lines and operated by one of multiple entities. Reliable and secure operation of this electric grid is of utmost importance for maintaining a country's economy and well-being of its citizens.

To operate the grid reliably and securely under all conditions, as well as adequately plan for the future, it is essential that one can robustly analyze the grid off-line and in real-time. At present, numerous analysis methods exist for operation and planning of the grid. These can be broadly categorized into one of the following: i) steady-state analysis in the frequency domain (power flow, three phase power flow, and harmonic analyses), ii) transient and steady-state analysis in time domain, iii) analysis for optimal dispatch of resources, and iv) other market dispatch-based analyses. Among these analyses, fundamental frequency based steady-state analysis (power flow and three-phase power flow) is essential for the day-to-day operation as well as future planning of the grid. Furthermore, the solution to the steady-state analysis serves as the initial state for transient analysis as well as the optimal power flow problem. Due to its critical importance, research has produced significant advances toward improving the convergence of these solution methods [5]-[12] .

At present, steady-state simulation is divided into two domains, high-voltage transmission systems and sub-station level voltage distribution systems. Disparate methods exist for analyzing these two (transmission and distribution) systems. The steady-state solution for the high voltage transmission system is obtained via positive sequence AC power flow analysis (often referred to as power flow analysis), whereas the steady-state operating point for the distribution system is obtained via three-phase AC power flow analysis. The industry standard for solving the positive sequence AC power flow problem is the 'PQV' formulation [1], wherein nonlinear power mismatch equations are solved for bus voltage magnitudes and angles that further define the steady-state operating point of the system. In contrast, the backward-forward sweep method [2] and the current injection method (CIM) [3] are primarily used for obtaining the steady-state solution of the three-phase power flow problem for the distribution grid.

In their existing forms, the solution methods for power flow and three-phase power can suffer from lack of convergence robustness [5], [10]. The 'PQV' based formulation for the positive sequence power flow problem is known to diverge or converge to non-physical solutions for ill-conditioned [2] and large scale (>50k buses) systems [20], where a non-physical solution corresponds to a system that contains low voltages or demonstrates angular instability. For distribution system problems, the backward-forward sweep method that was proposed to solve radial and weakly meshed distribution systems with high R/X ratio [2] has difficulties converging for heavily meshed systems with more than a single source of generation [12]. The CIM method based on Dommel's work in 1970 [4], like the equivalent circuit approach proposed in this paper, represents the currents and voltages in terms of rectangular coordinates, but is challenged by incorporation of multiple PV buses in the system [13]-[14]. In general, of the known challenges associated with convergence for existing power flow and three-phase power flow solution methods, the

This work was supported in part by the Defense Advanced Research Projects Agency (DARPA) under award no. FA8750-17-1-0059 for the RADICS program.
[1]Authors are with the Electrical and Computer Engineering Department, Carnegie Mellon University, Pittsburgh, PA 15213 USA, (e-mail: {amritanp, mjeremin, mwagner1, dbromber, pileggi}@andrew.cmu.edu).
[2]Author is with the Power System Laboratory, ETH Zurich, (e-mail: hug@eeh.ee.ethz.ch).
Digital Object Identifier 10.1109/TPWRS.2018.2863042

two that are most detrimental are convergence to low-voltage or unacceptable solutions [15] and divergence [5].

The objective and contribution of the approaches described in this paper is to provide *robust* power flow and three-phase power flow convergence. Specifically, a generalized approach for power flow and three-phase power flow analyses that can ensure convergence to correct physical solution independent of the choice of initial conditions.

The factors that are the most fundamental toward making these problems challenging are the use of non-physical representations for modeling the power grid components, and in the case of the 'PQV' formulation, the use of inherently non-linear power mismatch equations to formulate the problem. The non-physical representations of the system equipment may not capture the true behavior of the model in the entire range of system operation. For example, an approximated macro-model of a generator that is represented via positive sequence or three-phase PV node can result in convergence to a low-voltage solution or divergence due to its quadratic voltage characteristics. Similarly, the inherent non-linearities in the 'PQV' formulation almost always cause divergence for large (>50k) and ill-conditioned test cases [20] when solved from an arbitrary set of initial conditions. This lack of a physics-based formulation, along with the methods that can constrain the non-physics based models in their physical space, is what renders the existing power flow and three-phase power flow problem and solution approaches to be "non-robust."

To develop a robust solver for the steady-state solution of the power grid, it is imperative that the solver can efficiently and effectively navigate through the aforementioned challenges while converging to a solution that is both meaningful and correct. Intuitively and physically, both the transmission and distribution electric grids correspond to an electric circuit. Our approach toward solving the power flow and three-phase power flow problems is to utilize circuit modeling and formalism to develop new algorithms that will robustly solve them. Toward this goal, we propose a two-pronged approach. First, the use of an equivalent circuit formulation in terms of the true state variables of currents and voltages [16]-[18] to model both the transmission and distribution power grid (Sect. III.). Secondly, the adaptation and application of circuit simulation methods [19]-[22] to ensure robust convergence to correct physical solutions (Sect. IV.) for power flow and three-phase power flow problems. To demonstrate the interaction between the two, Sect. V of this paper discusses the general algorithm for solving the power flow and three-phase power flow problems. Several examples are shown which demonstrate the efficacy of our approach.

## II. Background

A power grid in its simplest form can be represented by a set of $\mathcal{N}$ buses, where the sets of generators $\mathcal{G}$ and load demands $\mathcal{L}$ are subsets of $\mathcal{N}$, which are further connected by a set of line elements, $\mathcal{T}_X$ and set of transformers $xfmrs$. Furthermore, there is a set of slack buses (one for each island in the system) represented by ξ. In addition to these, the power grid may contain other elements, such as shunts, flexible alternating current transmission system (FACTS), etc. The objective of steady-state analysis of the power grid is to model the fundamental frequency component of the power grid and solve for the complex voltages at its buses. The high voltage transmission network of the grid generally operates under balanced conditions, and therefore, the steady-state solution of the transmission network is obtained via positive sequence power flow model and analysis. In contrast, the distribution network of the power grid can operate under unbalanced conditions, and therefore we must apply three-phase power flow model and analysis to find the steady-state solution of the distribution grid. In the following sub-sections, we discuss the current formulations used for steady-state analysis of transmission and distribution networks and highlight their limitations.

### A. 'PQV' based Formulation for Positive Sequence Power Flow Problem

The 'PQV' based power flow formulation is the industry standard for solving for the steady-state solution of the high voltage transmission network. In this formulation, a set of $2(\mathcal{N} - \xi) - \mathcal{G}$ power mismatch equations are solved for unknown complex voltage magnitudes and angles of the system using the Newton Raphson (NR) method. The set of power mismatch equations are defined as follows:

$$P_G^i - P_L^i = |V_i| \sum_{l=1}^{\mathcal{N}} |V_l| (G_{il}^Y \cos \delta_{il} + B_{il}^Y \sin \delta_{il}) \quad (1)$$

$$Q_G^i - Q_L^i = |V_i| \sum_{l=1}^{\mathcal{N}} |V_l| (G_{il}^Y \sin \delta_{il} - B_{il}^Y \cos \delta_{il}) \quad (2)$$

where, $P_G^i + jQ_G^i$ and $P_L^i + jQ_L^i$ are the complex generation and complex load at the node $i$ and $G_{il}^Y + jB_{il}^Y$ is the complex admittance between the nodes $i$ and $l$.

In order to solve for unknown complex voltages $V_i \angle \delta_i$ in the system, the real and reactive power mismatch equations given by (1)-(2) are solved for the set of $(\mathcal{N} - \xi - \mathcal{G})$ buses in the system, whereas only real mismatch equations (1) are solved for the set of buses with generators $\mathcal{G}$ connected to it.

Importantly, this 'PQV' formulation is inherently non-linear, since the set of network constraints result in non-linear power mismatch equations independent of physics of the models used. For example, in the 'PQV' formulation, a linear network consisting of linear models for the slack bus, the transmission lines and the loads would correspond to a non-linear set of power mismatch equations, a feature that could result in convergence difficulties for systems even trivial in size.

### B. Current Injection Method for Three-Phase Power Flow Problem

Until recently, the backward forward sweep method was the most commonly used method for the steady-state analysis of the radial and weakly meshed distribution systems [2]. The method was preferred over the 'PQV' formulation due to the radial nature of the distribution grid and high R/X ratios of the distribution lines, both of which are known to cause convergence difficulties for the NR method [2] with 'PQV' formulation. However, the backward forward sweep method itself is prone to convergence difficulties for systems that are highly meshed or have multiple sources [12].

The current injection method (CIM) for the three-phase power flow problem [3] was proposed to address challenges associated with the 'PQV' formulation and the backward-forward sweep method. In the CIM formulation, the non-linear current mismatch equations for the system buses are solved via NR method for each individual phase with complex rectangular real and imaginary voltages ($V_{Ri}^{\Omega} + jV_{Ii}^{\Omega}$) as the unknown variables. The current mismatch equations for the three-phase power flow problem are defined as follows [3]:

$$\Delta I_{Ri}^{\Omega} = \frac{(P_i^{sp})^{\Omega} V_{Ri}^{\Omega} + (Q_i^{sp})^{\Omega} V_{Ii}^{\Omega}}{(V_{Ri}^{\Omega})^2 + (V_{Ii}^{\Omega})^2} \\ - \sum_{l=1}^{\mathcal{N}} \sum_{t \in \Omega_{set}} (G_{il}^{\Omega t} V_{Ri}^{t} - B_{il}^{\Omega t} V_{Ii}^{t}) \quad (3)$$

$$\Delta I_{Ii}^{\Omega} = \frac{(P_i^{sp})^{\Omega} V_{Ii}^{\Omega} - (Q_i^{sp})^{\Omega} V_{Ri}^{\Omega}}{(V_{Ri}^{\Omega})^2 + (V_{Ii}^{\Omega})^2} \\ - \sum_{l=1}^{\mathcal{N}} \sum_{t \in \Omega_{set}} (G_{il}^{\Omega t} V_{Ii}^{t} - B_{il}^{\Omega t} V_{Ri}^{t}) \quad (4)$$

where $\Delta I_{Ri}^{\Omega} + j\Delta I_{Ii}^{\Omega}$ is the net current mismatch in phase $\Omega$ at node $i$ and $(P_i^{sp})^{\Omega} + j(Q_i^{sp})^{\Omega}$ is the specified complex power injection at node $i$. The set $\Omega_{set}$ includes phases a, b and c.

Although, the CIM method is known to improve the convergence properties for heavily and weakly meshed three-phase radial distribution systems with high R/X ratio, the method is known to diverge for test-cases with high penetration of PV buses [13]. Traditionally, the number of PV buses in the distribution system were limited to a small number; however, with the advent of large scale installation of distributed generation (DGs) and voltage control devices in the distribution system this is no longer true. Therefore, it is essential that a robust three-phase power flow formulation can robustly handle high penetration of PV buses and other voltage control devices in the system.

## III. EQUIVALENT CIRCUIT FORMULATION

We extend the equivalent circuit approach in [16]-[20] for steady-state analysis of the transmission and distribution power grid to tackle the challenges exhibited by the existing formulations. This approach for generalized modeling of the power system in steady-state (i.e. power flow and three-phase power flow) represents both the transmission and distribution power grid elements in terms of equivalent circuit elements without loss of generality. It was shown that each of the power system components can be directly mapped to an equivalent circuit model based on the underlying relationship between current and voltage state variables. Importantly, this formulation can represent any physics based model or measurement based semi-empirical models as a sub-circuit, as shown in [24], [25] and [26], and these models can be combined hierarchically with other circuit abstractions to build larger aggregated models. In the following section, we discuss generic equivalent circuit representations of power system components for both the positive sequence power flow problem and the three-phase power flow problem. Note that throughout the paper, the symbol superscript $\Omega$ in the mathematical expressions of the equivalent circuit models represents a phase from the set $\Omega_{set}$ of three phases a, b and c for the three-phase problem and represents the *positive sequence* (*p*) component for the power flow problem.

### A. PV Bus or the Generator Model

In the equivalent circuit approach, the generator (PV) bus model is modeled via a complex current source [19] and has the same behavior as of the PV node in power flow and three-phase power flow problems. To enable the application of NR, this complex current source is split into real and imaginary current sources ($I_{RG}^{\Omega}$ and $I_{IG}^{\Omega}$, respectively). This is necessary due to the non-analyticity of complex conjugate functions [16]. The resulting equations for the PV model in the power flow and three-phase power problem are:

$$I_{RG}^{\Omega} = \frac{P_G^{\Omega} V_{RG}^{\Omega} + Q_G^{\Omega} V_{IG}^{\Omega}}{(V_{RG}^{\Omega})^2 + (V_{IG}^{\Omega})^2} \quad (5)$$

$$I_{IG}^{\Omega} = \frac{P_G^{\Omega} V_{IG}^{\Omega} - Q_G^{\Omega} V_{RG}^{\Omega}}{(V_{RG}^{\Omega})^2 + (V_{IG}^{\Omega})^2} \quad (6)$$

Additional constraints that allow the generators to control the voltage magnitude either at its own node or any other remote node in the system are modeled by a control circuits, as shown in the following subsection. In the case of power flow problem, a single control circuit is needed whereas for the three-phase power flow problem, three such control circuits are needed for each PV bus in the system. The reactive power $Q_G^{\Omega}$ of the generator acts as the additional unknown variable for the additional constraint that is introduced due to voltage control. In case of three-phase power flow, three such additional variables and constraints are introduced.

As an example, the equivalent circuit for the positive sequence model for a PV bus used in power flow is shown in Fig. 1 for the $k + 1^{th}$ iteration of NR. It is constructed by linearizing the set of equations (5)-(6) for the positive sequence parameters and then representing the resulting equations using fundamental circuit elements (detailed procedure for this provided in [16]). To construct the PV bus equivalent circuit for three-phase power flow problem, three such circuits are first constructed and then are connected in grounded-wye configuration.

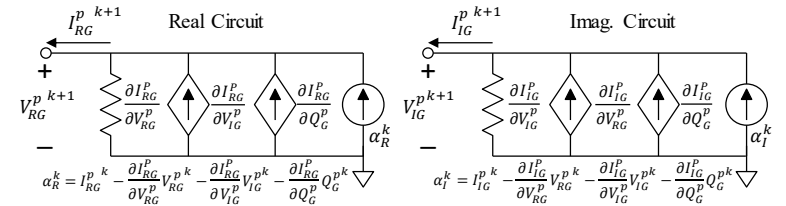

*Figure 1: Equivalent Circuit Model for PV generator model.*

### B. Voltage Regulation of a Bus

Numerous power grid elements such as generators, FACTS devices, transformers, shunts etc., can control a voltage magnitude at a given node in the system. Moreover, they can control the voltage magnitude ($V_{set}^{\Omega}$) at either their own node $\mathcal{O}$ or a remote node $\mathcal{W}$ in the system. In equivalent circuit formulation, we represent the control of the voltage magnitude by a control circuit (Fig. 2), which is governed by the following expression:

$$F_{\mathcal{W}}^{\Omega} \equiv (V_{set}^{\Omega})^2 - (V_{R\mathcal{W}}^{\Omega})^2 - (V_{I\mathcal{W}}^{\Omega})^2 = 0 \quad (7)$$

The circuit in Fig. 2 is derived from the linearized version of (7). For the power flow problem, it is stamped (i.e. values are added to the Jacobian in a modular way) for each node $\mathcal{W}$ in the system whose voltage is being controlled such that there exists at least one single path between the node $\mathcal{W}$ and the equipment's node $\mathcal{O}$ that is controlling it. Similarly, for three-phase power flow three of these circuits are stamped for each node $\mathcal{W}$. The additional unknown variables for these additional constraints are dependent on the power system device that is controlling the voltage magnitude. For example, the additional unknown variable for a generator is its reactive power $Q^\Omega$, whereas in the case of transformers, it is the transformer tap $tr^\Omega$, and for FACTS devices it is the firing angle $\varphi^\Omega$. The previous section already described how the additional unknown variable $Q^\Omega$ for PV buses is integrated into the respective equivalent circuits for generators.

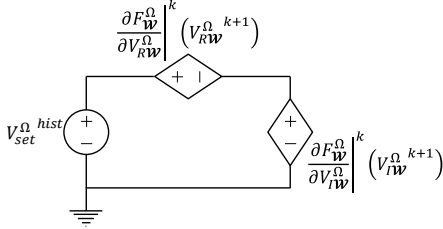

*Figure 2: Voltage magnitude constraint control equivalent circuit.*

### C. ZIP Load Model

In this section, we derive the positive sequence and three-phase model for the ZIP load. The ZIP load models the aggregated load in the system as a mix of constant impedance $(Z_P^\Omega + jZ_Q^\Omega)$, constant current $(I_P^\Omega + jI_Q^\Omega)$, and constant power $(S_P^\Omega + jS_Q^\Omega)$ load behavior, which can be mathematically represented as follows:

$$(P_i^{ZIP})^\Omega = Z_P^\Omega (|V_i^\Omega|)^2 + I_P^\Omega (|V_i^\Omega|) + S_P^\Omega \quad (8)$$

$$(Q_i^{ZIP})^\Omega = Z_Q^\Omega (|V_i^\Omega|)^2 + I_Q^\Omega (|V_i^\Omega|) + S_Q^\Omega \quad (9)$$

In the equivalent circuit approach, the equations for the ZIP load model can be re-written as:

$$(I_{Ri}^{ZIP})^\Omega = Y_P^\Omega V_{Ri}^\Omega - Y_Q^\Omega V_{Ii}^\Omega + \frac{S_P^\Omega V_{Ri}^\Omega + S_Q^\Omega V_{Ii}^\Omega}{(V_{Ri}^\Omega)^2 + (V_{Ii}^\Omega)^2} + \left(\sqrt{I_P^{\Omega\,2} + I_Q^{\Omega\,2}}\right) \cdot \cos(\delta_i^\Omega - I_{pf}^\Omega) \quad (10)$$

$$(I_{Ii}^{ZIP})^\Omega = Y_P^\Omega V_{Ii}^\Omega + Y_Q^\Omega V_{Ri}^\Omega + \frac{S_P^\Omega V_{Ii}^\Omega - S_Q^\Omega V_{Ri}^\Omega}{(V_{Ri}^\Omega)^2 + (V_{Ii}^\Omega)^2} + \left(\sqrt{I_P^{\Omega\,2} + I_Q^{\Omega\,2}}\right) \cdot \sin(\delta_i^\Omega - I_{pf}^\Omega) \quad (11)$$

where:

$$I_{pf}^\Omega = \tan^{-1}\left(\frac{I_Q^\Omega}{I_P^\Omega}\right) \quad (12)$$

$$\delta_i^\Omega = \tan^{-1}\left(\frac{V_{Ii}^\Omega}{V_{Ri}^\Omega}\right) \quad (13)$$

$$Y_P^\Omega + jY_Q^\Omega = \frac{1}{Z_P^\Omega + jZ_Q^\Omega} \quad (14)$$

For the load model given in (10) through (14), the constant impedance part of the load is linear, whereas the constant current and constant power part of the aggregated load is nonlinear. Once, (10)-(11) are linearized, they are used to construct the equivalent circuit models for both the power flow and three-phase power flow problem. The constructed three-phase model of the ZIP load model can either be connected in wye or delta formation. As an example, ZIP load model connected in wye and delta formation is shown in Fig. 3.

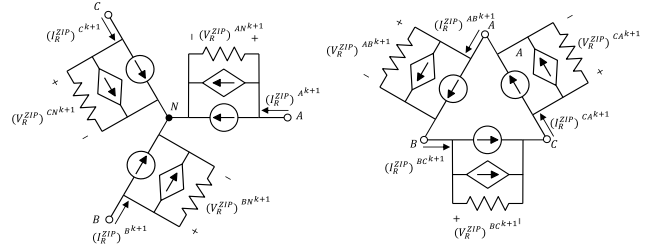

*Figure 3: Real circuit for a) wye connected ZIP Load Model (on left) b) delta (D) connected ZIP load model (on right).*

It is important to note that both the ZIP and PQ load models result in non-linear network constraints for both the 'PQV' and CIM formulations. In the 'PQV' formulation the non-linearities in the network constraints are due to the use of power mismatch equations whereas in the CIM, the non-linearities are due to PQ and ZIP model equations. These added non-linearities are one of the primary causes of divergence and convergence to low voltage solutions. To address this problem, we have proposed an accurate and yet linear *BIG* load model [25]-[27].

### D. BIG Linear Load Model

The *BIG* aggregated load model was proposed based on the circuit theoretic approach in [25]-[27] and aims to create a linear load model that can capture the true measure and sensitivity of the aggregated load in the system. The model is comprised of a susceptance (B), independent current source (I), and conductance (G). The complex governing equation of the generalized load current for the *BIG* load model is given by:

$$(I_R^{BIG})^\Omega + j(I_I^{BIG})^\Omega = (\alpha_R^{BIG})^\Omega + j(\alpha_I^{BIG})^\Omega + \left((V_R^{BIG})^\Omega + j(V_I^{BIG})^\Omega\right)((G^{BIG})^\Omega + j(B^{BIG})^\Omega) \quad (15)$$

where $(\alpha_R^{BIG})^\Omega + j(\alpha_I^{BIG})^\Omega$ represents the base value for the modeled aggregated load and the corresponding complex admittance $((G^{BIG})^\Omega + j(B^{BIG})^\Omega)$ captures the voltage sensitivities. For instance, a negative conductance in conjunction with complex current $\left((\alpha_R^{BIG})^\Omega + j(\alpha_I^{BIG})^\Omega\right)$ mimics the inverse current/voltage sensitivity relationship, similar to constant power (PQ) load behavior and positive conductance in conjunction with complex current source will represent the positively correlated current/voltage sensitivity relationship, similar to the impedance load behavior. Both the positive and negative impedances capture the change in load with voltage with respect to the portion of the load that is modeled by the current source. Fig. 4 shows the positive sequence (p) *BIG* load model. Similar to the ZIP load model, the three-phase *BIG* load model can be constructed by connecting the equivalent circuits of individual phases in wye or delta formation.

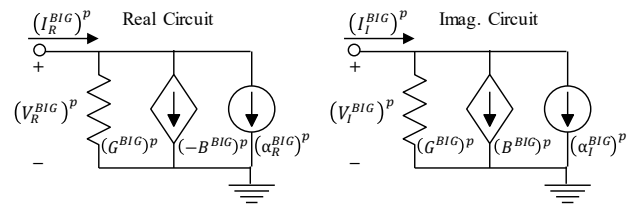

*Figure 4: Equivalent circuit of a BIG load model*

## IV. Circuit Simulation Methods

Decades of research in circuit simulation have demonstrated that circuit simulation methods can be applied for determining the DC state of highly non-linear circuits using NR. These techniques have been shown to make NR robust and practical for large-scale circuit problems [21]-[22] consisting of billions of nodes. Most notable is the ability to guarantee convergence to the correct physical solution (i.e. global convergence) and the capability of finding multiple operating points [28]. We have previously proposed analogous techniques for ensuring convergence to the correct physical solution for the positive sequence power flow problem [19]-[20]. In this section, we extend these methods to be used with positive-sequence power flow and three-phase power flow problems alike.

### A. General Methods

#### 1) Variable Limiting

The solution space of the system node voltages in a power flow and three-phase power flow problem are well defined. While solving these problems, a large NR step may step out of this solution space and result in either divergence or convergence to a non-physical solution. It is, therefore, important to limit the NR step before an invalid step out of the solution space is made. In [19], we proposed the variable limiting method to achieve the postulated goal for power flow problem. In this technique, the state variables that are most sensitive to initial guesses are damped when the NR algorithm takes a large step out of the pre-defined solution space. Note, however, that *not all* the system variables are damped for the variable limiting technique, as is done for traditional damped NR. Circuit simulation research has shown that damping the most sensitive variables provides superior convergence compared to damped NR in general [21].

To apply variable limiting in our prototype simulator for the power flow and three-phase power flow problem, the mathematical expressions for the PV nodes in the system are modified as follows:

$$I_{CG}^{\Omega\ k+1} = \varsigma \frac{\partial I_{CG}^{\Omega}}{\partial V_{RG}^{\Omega}} \underbrace{\left(V_{RG}^{\Omega\ k+1} - V_{RG}^{\Omega\ k}\right)}_{\Delta V_{RG}^{\Omega}} + I_{CG}^{\Omega\ k} \\ + \varsigma \frac{\partial I_{CG}^{\Omega}}{\partial V_{IG}^{\Omega}} \underbrace{\left(V_{IG}^{\Omega\ k+1} - V_{IG}^{\Omega\ k}\right)}_{\Delta V_{IG}^{\Omega}} + \frac{\partial I_{CG}^{\Omega}}{\partial Q_G^{\Omega}}\left(Q_G^{\Omega\ k+1} - Q_G^{\Omega\ k}\right) \quad (16)$$

where, $0 \leq \varsigma \leq 1$. The magnitude of $\varsigma$ is dynamically varied through heuristics such that convergence to the correct physical solution is achieved in the most efficient manner. The heuristics depend on the largest delta voltage ($\Delta V_{RG}^{\Omega}$, $\Delta V_{IG}^{\Omega}$) step during subsequent NR iterations. If during subsequent NR iterations, a large step ($\Delta V_{RG}^{\Omega}$, $\Delta V_{IG}^{\Omega}$) is encountered, then the factor $\varsigma$ is decreased. The factor $\varsigma$ is scaled back up if consecutive NR steps result in monotonically decreasing absolute values for the largest error.

#### 2) Voltage Limiting

An equally simple, yet effective, technique is to limit the absolute value of the delta step that the real and imaginary voltage vectors can make during each NR iteration. This is analogous to the voltage limiting technique used for diodes in circuit simulation, wherein the maximum allowable voltage step during NR is limited to twice the thermal voltage of the diode [22]. Similarly, for the power flow and three-phase power flow analyses, a hard limit is enforced on the normalized real and imaginary voltages in the system. The mathematical implementation of voltage limiting in our formulation is as follows:

$$(V_C^{\Omega})^{k+1} = \min_{V_C^{min}} \max_{V_C^{max}} \left((V_C^{\Omega})^k + \delta_S \min(|\Delta(V_C^{\Omega})^k|, \Delta V_C^{max})\right) \quad (17)$$

$$\min_{V_C^{min}} \max_{V_C^{max}} = \begin{cases} V_C^{max}, & \text{if } x > V_C^{max} \\ V_C^{min}, & \text{if } x < V_C^{min} \\ x, & \text{otherwise} \end{cases} \quad (18)$$

and $\delta_S = sign\left(\Delta(V_C^{\Omega})^k\right)$ and $C \in \{R, I\}$ represents the placeholder for real and imaginary parts.

Analogously, other system variables such as the reactive power $Q_G$ of the PV buses, can be limited by limiting the calculated currents $I_C^{\Omega} + \Delta(I_C^{\Omega})^k$ at NR step $k+1$ and then finding the new $Q_G^{k+1}$ from inverse function ($f^{-1}$) of limited $\left(I_C^{\Omega} + \overline{\Delta(I_C^{\Omega})^k}\right)$.

### B. Homotopy Methods

Limiting methods may fail to ensure convergence for certain ill-conditioned and large test systems when solved from an arbitrary set of initial guesses. To ensure convergence for these network models to the correct physical solutions independent of the choice of initial conditions, we propose the use of homotopy methods. Homotopy methods in past have been used to study the voltage collapse of a given network or to determine maximum available transfer capability [8]-[9]. They have also been researched for locating all solutions to a power flow problem [11], [30]. However, their usage for enabling convergence for hard to solve positive sequence and three-phase power flow problems has been limited at best. Of the proposed methods for better convergence [5], [23], most have suffered from convergence to low voltage solutions or divergence. On the other hand, some of them have been developed for formulations that are not standard for both positive sequence as well as three-phase power flow [6]. Furthermore, none of the previously proposed homotopy methods are known to scale up to test systems that are of the scale of the European or the US grids and in general are not extendable to the three-phase power flow problem.

In the homotopy approach, the original problem is replaced with a set of sub-problems that are sequentially solved. The set of sub-problems exhibit certain properties, namely, the first sub-problem has a trivial solution and each incrementally subsequent problem has a solution very close to the solution of the prior sub-problem. Mathematically this can be described via the following expression:

$$\mathcal{H}(x,\lambda) = (1-\lambda)\mathcal{F}(x) + \lambda\mathcal{G}(x) \quad (19)$$

where $\lambda \in [0, 1]$.

The method begins by replacing the original problem $\mathcal{F}(x) = 0$ with $\mathcal{H}(x, \lambda) = 0$. The equation set $\mathcal{G}(x)$ is a representation of the system that has a trivial solution. The homotopy factor $\lambda$ has the value of 1 for the first sub-problem and therefore the initial solution is equal to trivial solution of $\mathcal{G}(x)$. For the final sub-problem that corresponds to the original problem, the homotopy factor $\lambda$ has the value of zero. To generate sequential sub-problems, the homotopy factor is dynamically decreased in small steps until it has reached the value of zero.

In this paper, we discuss two homotopy methods that are specifically developed for the power flow and three-phase power flow analyses:

*1) Tx Stepping*

We proposed the "Tx Stepping" method in [20] specifically for the power flow problem. In this section, the method is further extended for the three-phase power flow problem.

   *a) General Approach*

In Tx stepping method, the series elements in the system (transmission lines, transformers etc.) are first "virtually" shorted to solve the initial problem that has a trivial solution. Specifically, a large conductance ($\gg G_{il}$) and a large susceptance ($\gg B_{il}$) are added in parallel to each transmission line and transformer model in the system. In case of three-phase power flow, a large self-impedance ($\gg Y_{\Omega\Omega}^{il}$) is added in parallel to each phase of the transmission line and transformer model. Furthermore, the shunts in the system, are open-circuited by modifying the original shunt conductance and susceptance values. Importantly, the solution to this initial problem results in high system voltages (magnitudes), as they are essentially driven by the slack bus complex voltages and the PV bus voltage magnitudes due to the low voltage drops in the lines and transformers (as expected with virtually shorted systems). Similarly, the solution for the bus voltage angles will lie within an $\in$-small radius around the slack bus angle. Subsequently, like other continuation methods, the formulated system problem is then gradually relaxed to represent the original system by taking small increment steps of the homotopy factor ($\lambda$) until convergence to the solution of the original problem is achieved. Mathematically, the line and transformer impedances during homotopy for the power flow is expressed by:

$$\forall il \in \{\mathcal{T}_X, xfmrs\}: \hat{G}_{il} + j\hat{B}_{il} = (G_{il} + jB_{il})(1 + \lambda\gamma) \quad (20)$$

and for the three-phase problem:

$$\begin{bmatrix} \hat{Y}_{aa}^{il} & \hat{Y}_{ab}^{il} & \hat{Y}_{ac}^{il} \\ \hat{Y}_{ba}^{il} & \hat{Y}_{bb}^{il} & \hat{Y}_{bc}^{il} \\ \hat{Y}_{ca}^{il} & \hat{Y}_{cb}^{il} & \hat{Y}_{cc}^{il} \end{bmatrix} = \begin{bmatrix} Y_{aa}^{il}(1+\gamma\lambda) & Y_{ab}^{il} & Y_{ac}^{il} \\ Y_{ba}^{il} & Y_{bb}^{il}(1+\gamma\lambda) & Y_{bc}^{il} \\ Y_{ca}^{il} & Y_{cb}^{il} & Y_{cc}^{il}(1+\gamma\lambda) \end{bmatrix} \quad (21)$$

where, $G_{il}$, $B_{il}$, and $Y_{\Omega\Omega}^{il}$ are the original system impedances and $\hat{G}_{il}, \hat{B}_{il}$, and $\hat{Y}_{\Omega\Omega}^{il}$ are the system impedances used while iterating from the trivial problem to the original problem. The parameter $\gamma$ is used as a scaling factor for the conductances and susceptances. If the homotopy factor ($\lambda$) takes the value one, the system has a trivial solution and if its takes the value zero, the original system is represented.

Along with ensuring convergence for a problem, Tx stepping avoids the undesirable low voltage solutions for the positive sequence power flow and three-phase power flow problem since the initial problem results in a solution with high system voltages, and each subsequent step of the homotopy approach continues and deviates ever so slightly from this initial solution, thereby guaranteeing convergence to the high voltage solution for the original problem.

   *b) Handling of Transformer Phase Shifters and Taps*

To "virtually short" a power system, we must also account for transformer taps $tr^\Omega$ and phase shifting angles $\Theta^\Omega$. In a "virtually" shorted condition, all the nodes in the system must have complex voltages that are near the slack bus or PV bus complex voltages, which can be intuitively defined by a small epsilon norm ball around these voltages. Therefore, to achieve the following form, we must modify the transformer taps and phase shifter angles such that at $\lambda = 1$, their turns ratios and phase shift angles correspond to a magnitude of 1 pu and $0°$, respectively. Subsequently, the homotopy factor $\lambda$ is varied such that the original problem is solved with original transformer tap and phase shifter settings. This can be mathematically expressed as follows:

$$\forall i \in xfmrs: \widehat{tr}_i^\Omega = tr_i^\Omega + \lambda(1 - tr_i^\Omega) \quad (22)$$
$$\forall i \in xfmrs: \hat{\Theta}_i^\Omega = \Theta_i^\Omega - \lambda\Theta_i^\Omega \quad (23)$$

   *c) Handling of Voltage Control for Remote Buses*

To achieve a trivial solution during the first step of Tx stepping it is essential that we also handle remote voltage control appropriately. Remote voltage control refers to a device on node $\mathcal{O}$ in the system controlling the voltage of another node $\mathcal{W}$ in the system. This behavior is highly non-linear and if not handled correctly can result in divergence or convergence to low voltage solution. Existing commercial tools for power flow and three-phase power flow analyses have difficulties dealing with this problem and suffer from lack of robust convergence when modeling remote voltage control in general. With Tx stepping we can handle this problem efficiently and effectively. We first incorporate a "virtually short path" between the controlling node ($\mathcal{O}$) and the controlled node ($\mathcal{W}$) at $\lambda = 1$, such that the device at the controlling node can easily supply the current needed for node $\mathcal{W}$ to control its voltage. Then following the homotopy progression, we gradually relax the system such that additional line connecting the controlling node ($\mathcal{O}$) and controlled node ($\mathcal{W}$) is open at $\lambda = 0$.

   *d) Implementation of Tx Stepping in Equivalent Circuit Formulation*

Unlike traditional implementations of homotopy methods, in equivalent circuit formulation we do not directly modify the non-linear set of mathematical equations, but instead embed a homotopy factor in each of the equivalent circuit models for the power grid components. In doing so we allow for incorporation of any power system equipment into the Tx stepping approach within the equivalent circuit formulation framework, without loss of generality. Furthermore, we ensure, that the physics of the system is preserved while modifying it for the homotopy method. Fig. 5 demonstrates how the homotopy factor is embedded into the equivalent circuit of the transformer.

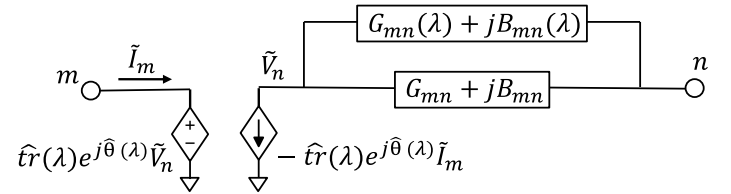

*Figure 5: Homotopy factor embedded in transformer equivalent circuit.*

*2) Dynamic Power Stepping*

Another homotopy technique that can ensure robust convergence for systems that have a low percentage of constant voltage nodes in the system is the dynamic power stepping method. Existing distribution systems tend to belong to these types of systems and therefore, dynamic power stepping can be applied to robustly obtain the steady-solution of the distribution grid by solving the three-phase power flow problem. This method has been previously described for the positive-sequence power flow problem in [19] and is analogous to the source stepping and gmin stepping approaches in standard circuit simulation solvers.

In the power stepping method, the system loads and generation are scaled back by a factor of $\beta$ until the convergence is achieved. If these loads and generations are scaled down all the way to zero, then the constraints for the PQ buses in the system result in linear network constraints. Similarly, current source non-linearities of the PV buses that are due to the constant real power are also eliminated. Therefore, by applying the power stepping factor, the non-linearities in the system are greatly eased and convergence is easily achieved. Upon convergence, the factor is gradually scaled back up to unity to solve the original problem. In this method, as in all continuation methods, the solution from the prior step is used as the initial condition for the next step. The mathematical representation of dynamic power stepping for the three-phase power flow and positive sequence power flow problem is as follows:

$$\forall G \in PV: \hat{P}_G^\Omega = \beta P_G^\Omega \quad (24)$$
$$\forall L \in PQ: \hat{P}_L^\Omega = \beta P_L^\Omega \text{ and } \hat{Q}_L^\Omega = \beta Q_L^\Omega \quad (25)$$

where, $PQ$ are all load nodes and $PV$ are all generator nodes.

## V. POWER FLOW AND THREE-PHASE POWER FLOW ALGORITHM

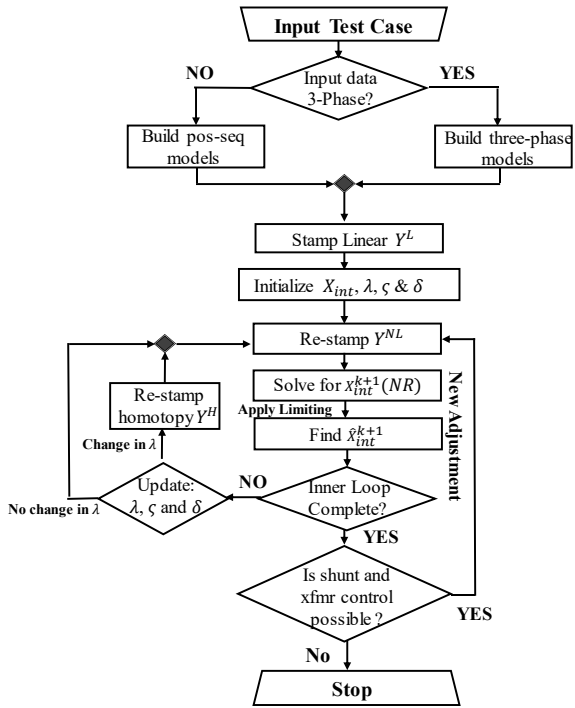

*Algorithm 1: Simulation algorithm for Positive Sequence and Three-Phase Power Flow Solver*

Algorithm 1 shows the recipe for the solving the positive-sequence as well as three-phase power flow problem in equivalent circuit approach with the use of circuit simulation methods. In this framework, the solver starts with building the system models based on the input file supplied. Linear models $(Y^L, J^L)$ are then stamped in the Jacobian matrix. Input state variables and other continuation parameters $(x_0, \delta, \varsigma, \lambda)$ are then initialized. Non-linear models are then stamped $(Y^{NL}, J^{NL})$ and NR is applied with limiting methods to calculate the next iterate for voltages and generator reactive powers $(\hat{X}^{k+1})$. Continuation and limiting parameters are then dynamically updated and homotopy models $(Y^H, J^H)$ are stamped or re-stamped if required to ensure convergence. Upon convergence of inner loop generator limits, switched shunts and transformer taps are adjusted and inner loop is repeated until final solution is achieved.

## VI. RESULTS

Example cases were simulated in our prototype solver SUGAR (Simulation with Unified Grid Analyses and Renewables) to demonstrate that the equivalent circuit approach along with circuit simulation techniques facilitates a robust framework for positive sequence power flow and the three-phase power flow analyses. The first set of results compare the solution of contingency analyses for two hard to solve cases with and without the use of circuit simulation methods to demonstrate the efficacy for these methods. All the further results compare the results of SUGAR (with circuit simulation methods) with other industry tools. The example cases for positive sequence power flow analyses include known ill-conditioned test cases and large network models that represent different operating and loading conditions for the eastern interconnection network of the US grid. For the three-phase power flow analysis, example cases include a set of standard distribution taxonomy cases [29], high density urban test cases [31], and a meshed transmission grid test case that was modified from a positive sequence to a three-phase network model. The results that follow demonstrate that the proposed framework along with the use of circuit simulation methods can ensure convergence to a correct physical solution for all the power flow and three-phase power flow cases, independent of the choice of the initial guess and thus overcomes the challenges faced by existing formulations.

### A. Circuit Simulation Methods

The purpose of following set of results is to demonstrate the robustness of the solver that is enabled due to the use of circuit simulation methods. To show this, contingencies were simulated on two (2) hard to solve test-cases that represent a real network for the subset of the US power grid. The base case for both simulations is first solved via Tx-stepping method and then used as an initial condition for the set of contingencies. The contingencies in the contingency set represent the loss of largest 10% of online generators and highest capacity lines and transformers dropped one at a time.

TABLE 1: COMPARISON OF SUGAR WITH AND WITHOUT CIRCUIT SIMULATION TECHNIQUES

| Case Id | # Bus | # Total Cases | SUGAR w/o Circuit Simulation Methods | | SUGAR with Circuit Simulation Methods | |
|---|---|---|---|---|---|---|
| | | | Converge | Diverge /Infeasible | Converge | Diverge /Infeasible |
| Case 1 | 5944 | 754 | 735 | 19 | 750 | 4 |
| Case 2 | 7029 | 801 | 706 | 95 | 793 | 8 |

The results in the Table 1 confirm that the circuit simulation methods when applied to equivalent circuit formulation can significantly increase the robustness of the power flow solver.

## B. Positive Sequence Power Flow Results

### 1) Ill-Conditioned and Large Test cases

A convergence sweep was run on the ill-conditioned 13659 bus PEGASE test case using the SUGAR solver and a standard commercial tool and their results were compared. Fig. 6 shows that SUGAR was able to robustly converge to the correct physical solution independent of the choice of the initial conditions, whereas the standard tool was highly sensitive to the choice of the initial guess and could converge to the correct physical solution only from a few samples for the initial guess.

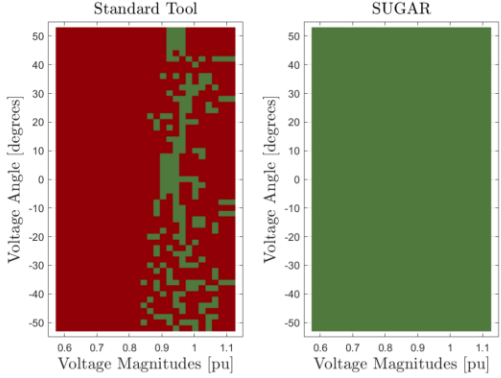

*Figure 6: Convergence sweep comparison for 13659 node PEGASE testcase between SUGAR and Standard tool. Red indicates divergence and green indicates convergence*

A similar convergence sweep was performed for larger test cases (> 75k+ nodes) that represent different loading and operating scenarios for eastern interconnection of the US grid. Simulations were performed on three different test cases for 15 different initial conditions each. Results are shown in Table 2. The set of initial conditions for all buses were identical and were uniformly sampled from:

$$V_{ang} \in [-40, 40], \ V_{mag} \in [0.9, 1.1]. \tag{26}$$

TABLE 2: CONVERGENCE PERFORMANCE FOR LARGE EASTERN INTERCONNECTION TEST CASES

| Case Name | # Nodes | Standard Tool | | SUGAR | |
|---|---|---|---|---|---|
| | | # Converge | # Diverge | # Converge | # Diverge |
| Case 1 | 80778 | 0 | 15 | 15 | 0 |
| Case 2 | 76228 | 0 | 15 | 15 | 0 |
| Case 3 | 81904 | 0 | 15 | 15 | 0 |

For the larger eastern interconnection test cases, the runtime per iteration is less than 0.4 seconds and is comparable to other simulation tools out in the market. The total computation time in general is dependent on the choice of initial conditions. A sufficiently close initial condition may result in convergence within 7 iterations whereas a totally random set of initial guesses may take up to 100 iterations with Tx stepping method.

### 2) Contingency Analysis

In the next set of results, we performed a set of contingency analyses with SUGAR and a standard commercial tool for two test cases that represent different network configuration of the eastern interconnection of the US grid. The initial guess for solving the contingency cases was chosen to be the operating point prior to the contingency. The set of contingencies in the experiment includes loss of generation ($\mathcal{L}_G$) and loss of branches ($\mathcal{L}_B$). The results are summarized in Table 3 and highlight the need for continuation methods to solve such problems robustly.

TABLE 3: COMPARISON OF CONTINGENCIES OF LARGE TEST CASES

| Case | # Nodes | Contingency* | Standard Tool | SUGAR |
|---|---|---|---|---|
| Case 1 | 76228 | $2\mathcal{L}_G$ | **Diverge** | **Converge** |
| | | $2\mathcal{L}_G + 2\mathcal{L}_B$ | **Diverge** | **Converge** |
| Case 2 | 78201 | $2\mathcal{L}_G$ | **Diverge** | **Converge** |
| | | $2\mathcal{L}_G + 2\mathcal{L}_B$ | **Diverge** | **Converge** |

*The number in front of $\mathcal{L}_G$ and $\mathcal{L}_B$ represents the equipment outage count. (For e.g. $2\mathcal{L}_G$ represents that two generators were lost during this contingency).

## C. Three-Phase Power Flow Results

### 1) Taxonomical Test Cases

Table 4 documents the results obtained from SUGAR three-phase solver for standard taxonomical cases and three large meshed test cases. The standard taxonomical cases include both balanced and unbalanced three-phase test cases. The first two meshed test cases are 342-Node Low Voltage Network Test Systems [31] that represent high density urban meshed low voltage networks. The third meshed test system is a high voltage 9241 node PEGASE transmission system that was extended to a balanced three-phase model. All these cases were simulated in SUGAR three-phase solver to validate its accuracy by comparing the obtained results against a standard distribution power flow tool GridLAB-D. Slight differences (less than 1e$^{-2}$) in the results were observed for cases between SUGAR and GridLAB-D and are due to default values used for unspecified parameters (e.g. neutral conductor resistance) in GridLAB-D.

TABLE 4: SUGAR THREE-PHASE RESULTS FOR TAXONOMICAL CASES

| Cases | #Nodes | Iter. Count | Deviation from GridLAB-D | |
|---|---|---|---|---|
| | | | Max. $\Delta V_{mag}$ [pu] | Max. $\Delta V_{ang}$ [°] |
| R1-12.47-1 | 2455 | 5 | 8.73E-04 | 9.94E-03 |
| R2-12.47-3 | 2311 | 5 | 6.56E-04 | 1.32E-02 |
| R3-12.47-3 | 7096 | 5 | 1.94E-03 | 3.89E-02 |
| R4-12.47-1 | 2157 | 5 | 6.81E-04 | 9.61E-03 |
| R5-12.47-5 | 2216 | 5 | 5.44E-05 | 4.20E-03 |
| NetworkModel 1 | 1420 | 3 | 3.38E-03 | 2.14E-03 |
| NetworkModel 2 | 1420 | 3 | 3.83E-03 | 6.00E-03 |
| case9241pegase* | 12528 | 5 | NA# | NA# |

* 9241 bus PEGASE transmission test case was extended to three-phase model
#The following case did not run in GridLAB-D

### 2) Ill-Conditioned Test Cases

To solve certain hard to solve ill-conditioned three-phase test cases, we made use of homotopy methods. To demonstrate one such example, we extended the standard 145 bus transmission system model into a balanced three-phase network model.

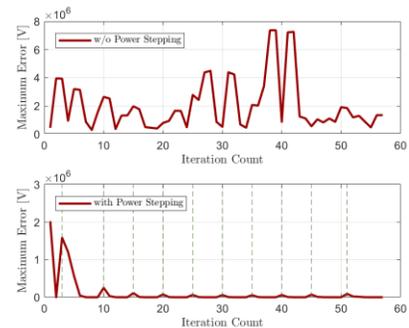

*Figure 7: Convergence of 145 bus test case for three-phase power flow with (bottom) and without (top) power stepping. For the power stepping case, the green dotted line represents the change in continuation factor λ*

Fig. 7 plots the convergence results for this test case with and without the use of dynamic power stepping. Without the use of dynamic power stepping, the test system did not converge within maximum number of allowable iterations; however, with the use of dynamic power stepping, the system robustly converged to correct physical solution.

VII. CONCLUSIONS

In this paper, we have demonstrated that the equivalent circuit approach with the use of novel circuit simulation methods can robustly solve for the steady-state solution of the transmission and distribution grid without loss of generality. This proposed formulation and the analogous circuit simulation methods can be generically applied to both the positive sequence power flow problem and the three-phase power flow problem. Importantly, our approach toward steady-state analyses of transmission and distribution grid ensures robust convergence to correct physical solutions, and in doing so enables robust contingency analyses, statistical analyses, and security constrained optimal power flow analyses. Furthermore, the proposed generic framework for transmission and distribution grid analyses can be extended for joint simulation of transmission and distribution circuits without loss of generality.

## IX. Biographies

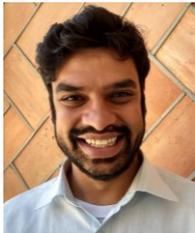

**Amritanshu Pandey** was born in Jabalpur, India. He received his M. Sc. Degree in electrical engineering from Carnegie Mellon University in Pittsburgh, PA in 2012. He is presently pursuing Ph.D. degree at Carnegie Mellon University. Prior to joining as a doctoral student at Carnegie Mellon University, he worked as an electrical engineer at MPR Associates Inc. from 2012 to 2015. He has previously interned at Pearl Street Technologies, ISO New-England and GE Global Research. His research interests include modeling and simulation, optimization and control of power systems.

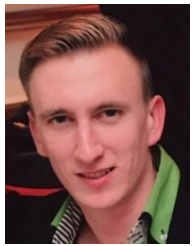

**Marko Jereminov** was born in Belgrade, Serbia. He received his B.Sc. Degree in Electrical Engineering from South Carolina State University, South Carolina, USA in 2016, and is currently pursuing Ph.D. degree in Electrical and Computer Engineering at Carnegie Mellon University, Pittsburgh, PA. He previously interned at Pearl Street Technologies, Pittsburgh, PA. His research interests include optimization, simulation and modeling of power systems.

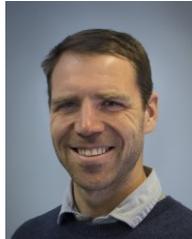

**Martin R. Wagner** received his B.Sc. degree in Electrical Engineering (2011) and M.Sc. degree in Microelectronics (2014) from the Vienna University of Technology in Vienna, Austria. His master thesis was created in collaboration with the Austrian Institute of Technology in Vienna, Austria. He is currently pursuing a PhD degree in Electrical Engineering at Carnegie Mellon University in Pittsburgh, PA, USA. His research interests include probabilistic methods applied to modeling and simulation of power systems.

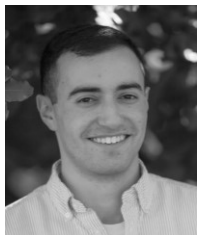

**David M. Bromberg** was born in Brooklyn, New York. He received the BS, MS, and Ph.D. degrees, all from Carnegie Mellon University in Pittsburgh, PA, in 2010, 2012, and 2014, respectively. After his graduate studies he joined Aurora Solar in Palo Alto, CA as a senior scientist, where he developed cloud simulation software for solar energy production modeling. In 2017 he co-founded Pearl Street Technologies, a power systems software company, and currently serves as its Chief Executive Officer. His research interests include robust, scalable algorithms for the simulation of electric power systems. He is a member of IEEE.

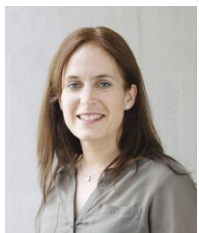

**Gabriela Hug** (S'05, M'08, SM'14) was born in Baden, Switzerland. She received the M.Sc. degree in electrical engineering and the Ph.D. degree from the Swiss Federal Institute of Technology (ETH), Zurich, Switzerland, in 2004 and 2008, respectively. After her Ph.D. degree, she was with the Special Studies Group of Hydro One in Toronto, Canada, and from 2009 to 2015, she was an Assistant Professor with Carnegie Mellon University, Pittsburgh, USA. She is currently an Associate Professor with the Power Systems Laboratory, ETH Zurich. Her research interest includes control and optimization of electric power systems.

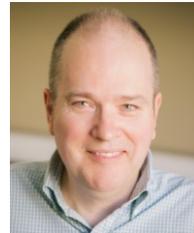

**Lawrence Pileggi** is the Tanoto professor of electrical and computer engineering at Carnegie Mellon University, and has previously held positions at Westinghouse Research and Development and the University of Texas at Austin. He received his Ph.D. in Electrical and Computer Engineering from Carnegie Mellon University in 1989. He has consulted for various semiconductor and EDA companies, and was co-founder of Fabbrix Inc., Extreme DA, and Pearl Street Technologies. His research interests include various aspects of digital and analog integrated circuit design and design methodologies, and simulation and modeling of electric power systems. He has received various awards, including Westinghouse corporation's highest engineering achievement award, the 2010 IEEE Circuits and Systems Society Mac Van Valkenburg Award, and the 2015 Semiconductor Industry Association (SIA) University Researcher Award. He is a co-author of "Electronic Circuit and System Simulation Methods," McGraw-Hill, 1995 and "IC Interconnect Analysis," Kluwer, 2002. He has published over 300 conference and journal papers and holds 39 U.S. patents. He is a fellow of IEEE.